# Accessing topological surface states and negative MR in sculpted nanowires of Bi$_2$Te$_3$ at ultra-low temperature

**Reena Yadav**[†,‡], **Biplab Bhattacharyya**[‡], **Animesh Pandey**[†,‡], **Mandeep Kaur**[‡], **R. P. Aloysius**,[†,‡] **Anurag Gupta**,[†,‡] and **Sudhir Husale**,[†,‡] *

[†] Academy of Scientific and Innovative Research (AcSIR), Ghaziabad-201002, India.

[‡] National Physical Laboratory, Council of Scientific and Industrial Research, Dr. K. S Krishnan Road, New Delhi-110012, India.

*E-mail: husalesc@nplindia.org

**Abstract**

Milling of 2D flakes is a simple method to fabricate nanomaterial of any desired shape and size. Inherently milling process can introduce the impurity or disorder which might show exotic quantum transport phenomenon when studied at the low temperature. Here we report temperature dependent weak antilocalization (WAL) effects in the sculpted nanowires of topological insulator in the presence of perpendicular magnetic field. The quadratic and linear magnetoconductivity (MC) curves at low temperature ($> 2K$) indicate the bulk contribution in the transport. A cusp feature in magnetoconductivity curves (positive magnetoresistance) at ultra low ($< 1K$) temperature and at magnetic field ($< 1T$) represent the WAL indicating the transport through surface states. The MC curves are discussed by using the 2D Hikami –Larkin-Nagaoka theory. The cross-over/interplay nature of positive and negative magnetoresistance observed in the MR curve at ultra-low temperature. Our results indicate that transport through topological surface states (TSS) in sculpted nanowires of Bi$_2$Te$_3$ can be achieved at mK range and linear MR observed at ~ 2 K could be the coexistence of electron transport through TSS and contribution from the bulk band.

Supplementary material for this article is available online

## 1. Introduction

Topological insulators (TI) possess very interesting transport through topologically protected metallic surface states (TSS)[1,2] without backscattering events and these TSS are potentially important for achieving the dissipation less electron transport in nano devices. Electron's spin and momentum are locked in these materials and are protected by the time reversal symmetry due to which electron acquires π Berry phase. Accessing transport through TSS is a challenging task because influence of the bulk channel which often gives signatures of the multiple conduction channels[3,4]. Dominance of bulk states indicates the fermi level in the bulk bands which impedes the transport through the TSS. WAL (weak anti-localization) effect is supposed to be a property of topological surface states and occurrence of it indicates the TSS dominated transport.

The magnetoresistance (MR) study is a simple method to investigate the type of conduction mechanism such as weak antilocalization (WAL) and weak localization (WL) effects which represent corrections to the conductivity arising from quantum interference of the electron wave functions between two time-reversed paths. These effects are noticeable in quantum diffusive regime where mean free path is much shorter than phase coherence length of quasi particles. Numerous studies have shown that these corrections in conductivity are affected by thickness of films, interaction of surface state and bulk carriers, presence of disorder and electron-electron interactions, where a crossover from WAL to WL can be observed[5].

The observance of linear MR (magnetoresistance) and negative MR (NMR), their cross over are important for technological applications as well understanding of these MR signatures is a basic research problem and recently it has been subject of intense research[6-14] in topological insulator based materials or in strongly correlated systems. NMR has been observed in non-magnetic materials such as Dirac and Weyl semimetals resulting from chiral anomaly[15,16]. However, in topological insulators NMR is not well defined and the observation of unusual NMR in these material is signature of peculiar physics[9,11,17]. Thus, there is growing interest to understand the mechanism for origin of NMR in TIs based nanomaterials.

Further robustness of the surface states is mainly theoretically predicated and there is very less work has been done on the experimental testing of TSS to answer very basic question whether one can demonstrate the robustness of TSS in the presence of introduced disorders (nonmagnetic impurities and deformations) in the sample without breaking the fundamental symmetry.

There are some strategies to make bulk insulating samples for accessing the transport through surface states e.g. suppressing bulk carriers by metal doping[18], changing the experimental conditions like gate voltage[19], improving the crystal quality[20], use of ultrathin samples[21], etc. The other alternative approach of nanosculpting or nanomilling reduces the size of sample, from micron sheets to nanowire and this method has been reported in the past[22-26]. The nanostructures such as nanowires or nanoribbons are more preferable over the bulk samples because of the reduction in surface to volume ratio which is useful for realizing the transport through the TSS. It has been observed that sculpting width of a nanowire down to 150 nm can be made easily and further reduction in the width of the nanowire may lose the crystalline nature or becomes more amorphous[26]. The Ga ions used during sculpting of the samples are also inherently introduces the impurity in the sample which might change the transport property at low temp transport[10]. Thus this method inherently introduces the deformations or impurities in the samples which is simple and important for investigating the robustness of TSS and MR studies.

Hence to overcome the dominance of bulk contribution, to study the MR signatures and to check the robustness of TSS we reduce the sample dimension by simply sculpting the nanosheeets into the nanowires or nanoribbons and we report the magnetoconductivity measurements on bismuth telluride nanowires at ultra low temp. The temp dependent evolution of WAL effects in magnetoresistance/conductivity curves have been studied here. The MR shows temp dependent evolution of three phases i.e. quadratic, linear and nonlinear. We analyze these results using Hikami –Larkin-Nagaoka (HLN) equations and study the phase coherence length, its temperature dependence and 2D nature of the transport. Surprisingly we find negative MR at ultra low temp < 0.12 K and crossover between positive MR to negative MR was observed which we correlates could be due to disorder in the system. Our results indicate that LMR (linear magnetoresitance) observed at high temp range (>2K) is not sufficient for accessing the TSS and measurements at ultra low temp give more access to TSS.

*2. Results and discussion*

Figure 1 shows the temperature dependent quadratic to linear MR change in the presence of perpendicular magnetic field. Inset in Fig. 1a shows the nanowire device used for the low temp measurements where sculpted $Bi_2Te_3$ nanowire is electrically contacted by using Pt electrodes. The MR curves are normalized and plotted for change in MR %. The curves measured at 10K and 5 K show quadratic or parabolic nature of MR with no linearity. At high temperature logarithmic MR mostly indicates the presence of Coulomb interaction. The quadratic nature of MR has been observed in different topological systems mainly indicating the dominance of bulk transport[27,28]. The curves at 2K and 1K represent linear relationship (range 0-8 T). Contrast to pure metals which show a quadratic MR at low temp and saturates at high magnetic field but materials like topological insulators display non saturating positive LMR. The non-saturating linear MR often described as an origin of all states separated by landau levels. At 1K, MR shows sharp dip at low field corresponding to the suppression of WL (weak localization) effect. In the literature, MR measurements showing WAL effects were heavily used to prove the transport through TSS[13,14,19,29-31].

The dependency of MR on $1/B^2$ was observed for high temp 5 K and 10 K and is shown in the Fig. 1a. The black line curves show the HLN fitting performed using following equation no. 1. HLN analysis helps in understanding the phase coherence length and strength of spin orbit coupling of the 2D system and rigorously used for the study of electron localization problems involving spin orbit couplings and the magnetic scattering. Strong SOC (spin orbit coupling) suppresses electron – electron interactions hence we find usefulness of HLN equation to explain the perpendicular field MR.

$$\frac{R(B)-R(0T)}{R(0T)} = -\left[\frac{\alpha e^2}{\pi h}\left\{\Psi\left(\frac{1}{2}+\frac{\hbar}{4eL^2B}\right) - \ln\left(\frac{\hbar}{4eL^2B}\right)\right\} + \beta B^2\right] * R(B) \qquad (2)$$

The logarithmic dependence of magnetoconductivity at high temp can be understood with the help of the electron -electron interaction which shows localization tendency indicating disordered nature. The arrows in Fig. 1 indicate that the equation 1 does not show good fit at low field (< 1T) and deviates at high field ∼ (> 8T) for the temp 2K and 1K. At 2K, it clearly shows the linear MR but cusp like feature at low magnetic field is not convincing which might indicate that both TSS and bulk bands contributes during the electron transport.

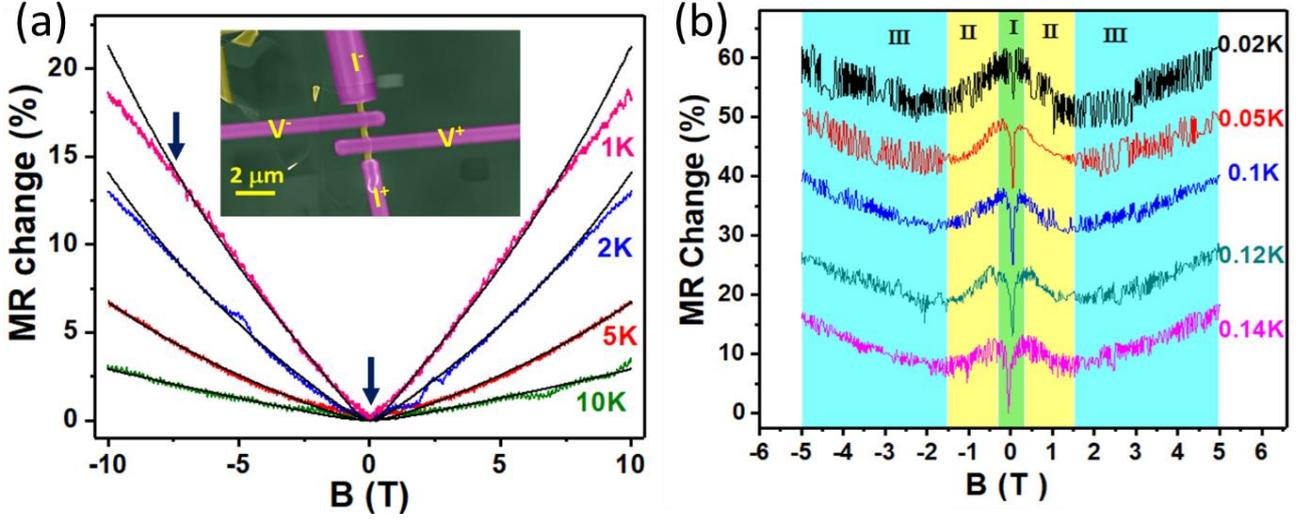

**Figure 1.** MR study of $Bi_2Te_3$ nanowire. a) MR measurements performed for temp $\geq$ 1K. Inset represent the device image. b) MR measurements performed at ultra low temp $\leq$ 0.14 K. Note that curves have been shifted for clarity

Fig. 1b shows the MR curves measured below 1K. Here we have observed three regions in MR curves namely occurrence of sharp cusp at very low fields which is represented by region I. WAL effects showing sharp cusp in MR curves have been observed in TI systems like bismuth selenide nanoribbons[32], $Bi_2Se_3$ thin films[33,34]. The origin of such WAL effects is reported as the negative interference in electrons path and commonly observed in TIs which we have explained in the next section with the help of HLN analysis (equation 2). Region II for applied field B from ~ 0.4 T to 1.5 T shows the negative MR. Recently negative MR study in Dirac metal or semimetal is very hot topic in research due to its different origin [10]. Interestingly at critical field ~1.6 T crossover takes place and curve displays positive MR as shown in the region III which we have explained in the later part.

$$\Delta G(B) = G(B) - G(0T) \cong \frac{\alpha e^2}{\pi h}\left\{\Psi\left(\frac{\hbar}{4eL^2B} + \frac{1}{2}\right) - ln\left(\frac{\hbar}{4eL^2B}\right)\right\} \qquad (2)$$

HLN relation describes the correction in the conductance in presence of spin orbit coupling to the temperature dependent WAL effects. In presence of perpendicular magnetic field, Zeeman Effect is suppressed in topological insulator due to strong spin orbit couplings[35]. HLN gives the quantum conductivity correction for both SOC and non SOC systems, for non SOC systems alpha prefator value changes sign and if in equation 1, alpha is set 0, the quantum correction term is no more and only classical MR is present. For the first part of equation, quantum term is valid at low field but when we add second classical parabolic term in equation 1 then it accounts for higher fields too. Further HLN fit estimate the phase coherence length, and a prefactor (fitting parameter) which is mostly reported negative for topological insulator systems and observed in the range -0.3 to - 1.1.

Fig. 2a shows the HLN fitting performed for temp range 1K to 0.3 K range. MC curves clearly show the formation of cusp at low field only (-0.5 to 0.5 T). Observance of WAL at low temp suggests that the self-intersecting scattering trajectories of electron follows destructive quantum interference. The conductivity of the nanowire at low temp increases this could be due to suppression of backscattering and decoherence of electrons. An increase in magnetic field destroys the WAL interference effects by lifting the time reversal symmetry. For higher fields B, the MC curves show linear dependency on the applied B for all the temp ranges as shown in the graph. The cusp becomes more sharp when measured for the temp range 0.17 K to 50 mK and interestingly the linearity observed at high field for the temp range 1 K to 0.3 K vanishes at this range (Fig 2b).

Fig. 2c shows the α prefactor that tells the mechanism of transport (the number of channels present in the interference effect). The variation of α from -1 to -0.5 indicate the presence of transport through TSS and bulk bands i.e. both are weakly coupled and bulk is still contributing in the conductance. The obtained fit values of prefactor α and phase coherence length are shown in the Fig. 2c. If the α value is 0 that corresponds to strong magnetic scattering and incoherent quantum phases. The α values about -0.5 suggests surface state dominated transport and observed in presence of strong SOC and weak magnetic scattering. In this regime destructive interference occurs which give rise to WAL. Since our fitted values of α are not exactly - 0.5 for all the temp that means the regime where WAL and WL are competing with each other. Note that experimentally it is very difficult to prove the transport through TSS with single mechanism only and hence we assume the value alpha indicates the intermediate regime where the surface state dominated transport is present.

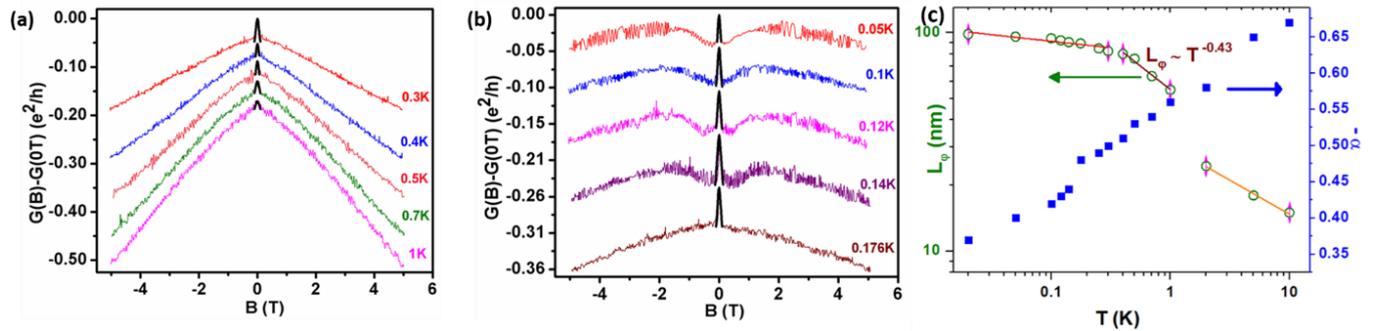

**Figure 2.** Magnetoconductivity (MC) measurement data and HLN analysis of $Bi_2Te_3$ nanowire. a) MC curves measured at ≥ 0.3K and HLN curve fit (black line) at low field. b) MC curves measured at ≤ 0.17K. Black line curve is the HLN fit. Note that curves are shifted for clarity. c) Represent the HLN fit parameter α and $L_\phi$ as a function of temp.

The phase coherence length ($L_\phi$) was obtained by using the above HLN fit equation 2 for the various temperature range as shown in the Fig. 2c which is complicated and represents contributions from several effects. We observed that $L_\phi$ decays with the increase in temperature and follows the power law fit relation $L_\phi \alpha T^P$. If the value of fit parameter P is close to -0.5 which indicates two dimensional nature of the transport[36] and dephasing due to prominent electron- electron interaction. The values less than -0.5 suggests the quasi one dimensional nature of the transport system. At higher temp, fit deviates which might indicate the electron phonon dominant interactions.

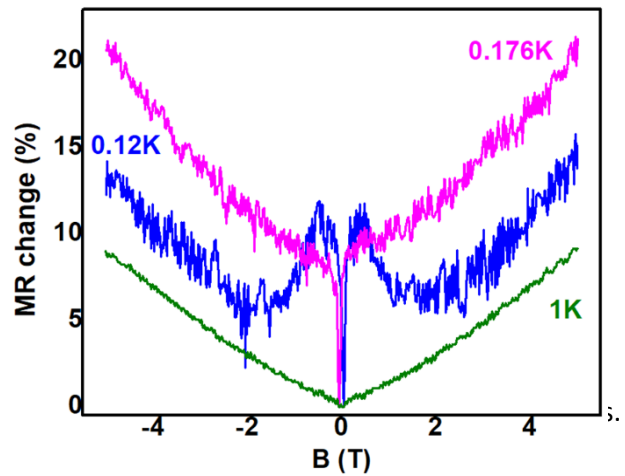

Figure 3 shows the comparison of three MR curves measured at 1K, 0.176 K and 0.12K. The transition in MR curves are clearly visible from linear, linear with sharp cusp and mixed behavior respectively. The strong spin orbit scattering is supposed to show the weak antilocalization effects. Spin orbit scattering can act like spin dependent magnetic scattering and this effect should overcome the transport through the bulk channels and may give rise the single channel representing the WAL and we have observed sharp cusp formation at temp far below (0.176 K) than the temp (1K) where we saw linear MR with no convincing WAL effect. Strikingly MR curve at 0.12 K shows negative MR for the field window (0.2 T to 2.5 T). The negative (positive) contribution in MR (MC) curves originate at 0.14 K. The positive slope in MR curve is a characteristic of the WAL effects and negative slope indicates the WL effects due to the presence of strong SOC. Understanding the origin of it cannot be explained by the HLN model used here. Positive LMR and MR with sharp cusp have been observed and very common in topological insulators because magnetic field localizes electron in cyclotron orbits while negative MR is uncommon effect which underlies interesting physics. NMR has been previously observed in non-magnetic materials such as Dirac and Weyl semimetals and which is reported due to chiral anomaly[15,16]. However, in topological insulators chiral anomaly is not well defined and the observed NMR has created a great confusion in the past[9,11,17]. MR switching from positive to negative MR in the ultrathin films of $Bi_2Se_3$ was also reported due to the interference of top and bottom surface wave functions which opens a gap in the surface states[37]. Previously negative MR was observed due to disorder formed in the sample which increases the density of defect states whereas the surface state were unaffected[11]. Other systems such as thin films of $(Bi_{1-x}Sb_x)_2Te_3$ and $Bi_2(Te_{0.4}Se_{0.6})_3$ have also reported NMR due to Anderson localization in disordered systems and WL effects[17,38].

Crossover of weak anti-localization to weak localization is possible if Berry phase = 0, in case of WAL Berry phase changes from $\pi$ to $2\pi$ and this could be the one reason to observe such change in the MR. Fermi level shifting causes competition between TSS electrons and bulk electrons and bulk gap can shift either valence bands or conduction bands. Thus Fermi level shift is also responsible for the crossover of the positive to negative magnetoconductivity. The WL is observed if the system is governed by the constructive interference of coherent quantum phases and often these type of systems show the presence of weak SOC or magnetic scattering. Previoulsy crossover from WAL to WL have been observed in many experiments, e.g. magnetically doped surface states[39], due to presence of bulk states in topological insulator[19] etc.

In our present result, we observed a combination of WAL and WL in MR data. At low temp, where topological surface states (TSS) are dominant, we observed a WAL cusp near 0T, which is suggestive of the non-trivial $\pi$-Berry phase that's a signature of TIs. However, with increasing temp, we observe that at 0T cusp becomes weaker and fades away, suggesting more bulk conduction channels in the sample due to higher thermal activation energies. Also, we observe that with increasing B-field, there is an interplay between WAL and WL features. Previously such effects have been characterized using spin-orbit scattering time ($\tau_{SO}$) and electron dephasing time ($\tau_\phi$)[40]. In regime 1 (as shown in the fig 1b), $\tau_\phi \gg \tau_{SO}$, this shows that spin orbit scattering leading to frequent spin flips is significant. The quantum interference of the electron wave functions between two time-reversed paths is destructive and manifests WAL effect (PMR). In regime 2 (fig 1b), $\tau_{SO} > \tau_\phi$, means spin orbit scattering is weak and it barely affects the spin orientation. Resistance of the material decreases with increasing magnetic field in this regime which results into WL (NMR). This intermediate regime is represented as a crossover from WAL to WL and vice-versa. For regime 3(fig 1b), $\tau_\phi > \tau_{SO}$, suggests strong spin orbit coupling leading to large increase in the resistance with magnetic field. This can be seen as large PMR in the devices. This physical picture can also be understood by using Berry phase language. Berry phase is a geometrical phase accumulated with a quantum mechanical system when system is subjected to an adiabatic cyclic process[41]. Time reversed scattering trajectories are equivalent to moving an electron on fermi surface by one cycle. So, electron picks up a Berry phase:

$$\phi_b \equiv -i \int_0^{2\pi} d\varphi \, \langle \psi k(r) | \frac{\partial}{\partial \varphi} | \psi k(r) \rangle = \pi (1 - \Delta/2E_F)$$

For massless Dirac fermions which lies on the surface states, $\Delta=0$ so $\phi_b = \pi$, which suggests quantum interference between time reversed trajectories is destructive that suppresses backscattering and enhances the conductivity, leading to occurrence of WAL effect. While for massive bulk electrons, where mass term is the band gap between conduction and valence band, so $\phi_b=0$. In this large-mass limit, quantum interference changes from destructive to constructive and crossover from WAL to WL takes place.

Note that here we used sculpted nanowires which could inherently introduce the disorder in the system. It has been observed that the locally conducting puddles of hole or electrons appear in some regions when the valence band top or conduction band bottom crosses the chemical potential. Such system coexist positively and negatively charged empty donors and occupied acceptors respectively. The thermally activated carriers do not show any significant screening at low temp and charged impurities pile up the fluctuations of the Coulomb potential[7].

In our previous work on FIB fabricated $Bi_2Te_3$ nanowires at ultra-low temp under in-plane parallel B-field, we demonstrated that spin-dependent scattering of surface and bulk electrons. The local magnetic moments produced by Ga ion milling contamination (with the help of first principles calculations) leads to the negative magnetoresistance in $Bi_2Te_3$ nanowires[10]. In this case under in-plane parallel B-field, Dirac cone only shifts in the X-Y plane with no gap term introduced in the energy eigenvalue $E = \pm\sqrt{(Ak_x)^2 + (Ak_y - m_x)^2}$, if B-field is in-plane-parallel in x-direction.

In present study, we have used out-of plane perpendicular B-field, where all such magnetic moments will align perpendicular to surface resulting into zero exchange interaction energy with surface electron spin (which is in-plane perpendicular to electron momentum direction). But, due to out-of plane perpendicular external B-field, time reversal symmetry will be broken, creating a "mass" or magnetic gap in the previously gapless linear Dirac spectrum. $E = \pm\sqrt{(Ak_x)^2 + (Ak_y)^2 + m_z^2}$, if B-field is out-of-plane perpendicular in z-direction, $k$ is the wave-vector and $m$ is the magnetic term. This will result into resistance increase (positive MR) with B-field.

It might be possible that not all magnetic moments align in perpendicular field direction and some in-plane components might exist. However, these few in-plane moments will be then randomly oriented and will cause spin scattering, which should not decrease resistance and show negative MR. Scattering should cause resistance to increase and MR to show positive slope. In our previous work under in-plane (parallel to current) B-field, we observed large negative MR even at low B-fields around zero. The magnetic moments (due to Ga contamination) alignment is perpendicular to TSS spin in such case, which reduces spin scattering and causes this negative MR. Therefore, in present work, we believe that small negative MR in region 2 at T <=

0.14 K observed in these nanowires under out-of plane perpendicular B-field can be attributed to the small bulk contribution, where Weak localization effect is dominant. Overall MR change from negative to positive value is a complex phenomenon and more experimental work is necessary for understanding the other phenomena involved e.g. fermi surface nesting arising from the quantum oscillations.

*3. Conclusion*

It is known that pure metals show a quadratic MR at low temp and saturates at high magnetic field but materials like topological insulators are mostly reported for non-saturating positive LMR hence it is important to study MR in TIs based nanodevices. We observed evolution of WAL effects from broad cusp to sharp cusp as a function of temp in sculpted nanowires of topological insulator. This indicates that bulk contribution to surface state transport can be minimized effectively at low temp. Strikingly, the negative magnetoresistance was also found at temp 0.14 mk range at low B field just after the sharp cusp and origin of it needs more investigations. Our results confirm that negative MR (NMR) can be observed even for field applied perpendicular and supports the fact that NMR is not limited to Weyl semimetals only. Here we have studied the WAL signatures in robust sculpted nanowires of $Bi_2Te_3$ and have analyzed the temp dependent WAL effects by using HLN equation. The important physical parameters α (related to SOC strength), and phase coherence length $L_\phi$ indicate the surface state dominated transport in the sculpted nanowire of TI which is considered as robust materials for back scattering events. Our results indicate the sculpted wires do show WAL effect of TSS and negative MR if measurements done at mk range and samples can be investigated further at such temp. Overall there is necessity of performing MR measurements at ultra low temp where exotic quantum nature dominates.

**Experimental**

**Fabrication procedure:** $Bi_2Te_3$ nanowires were fabricated from thin flakes of $Bi_2Te_3$ by using a scotch tape method and FIB milling technique as described in the literature [23,24]. The high quality $Bi_2Te_3$ (99.99 pure) material was purchased from Alfa Aesar Company. Initially, Si/SiO$_2$ chips (p type, highly doped with boron and resistivity ~ 0.001 – 0.005 Ω-cm) were ultrasonicated for 10 min and then cleaned with acetone, isopropanol, methanol and deionized water. The plasma cleaning with oxygen gas was performed for about 5 min. Au/Ti contact pads were deposited on cleaned wafers using DC magnetron sputtering. The standard scotch tape method was used for micromechanical cleavage of $Bi_2Te_3$ bulk crystals. The exfoliated flakes were deposited on Si/SiO$_2$ chips with predefined Au/Ti contacts. $Bi_2Te_3$ flakes with random size and arbitrary thickness were produced in this method, which were observed under optical microscope (Olympus MX51) and field emission scanning electron microscopy (FESEM, Zeiss–Auriga) and very thin flakes were further localized. The atomic force microscopy (AFM) and cross-sectional FESEM technique were employed to determine thickness of these localized thin flakes. A focused ion beam (Zeiss- Auriga) milling technique was used to fabricate the nanowire from located thin flake. Experimentally, it has been observed that FIB fabrication can introduce some $Ga^+$ ion contamination. Hence extra care was taken during the device fabrication and sample was never exposed to $Ga^+$ ion beam imaging. A low ion beam current (50 pA) was used to etch some portion of the deposited flake and beam exposure time was less than 20 sec. For electrical measurements, four-probe geometry with two voltage electrodes, two current electrodes were deposited on the $Bi_2Te_3$ nanowire using a gas injection system (GIS, FIB based metal deposition).

**Transport measurements:** The transport measurements of the nanowires at ultra-low temperatures were carried out in a Dilution Refrigerator (Triton-200, Oxford Instruments) using low frequency ac (17 Hz) lock-in technique in the four-terminal configuration. In order to filter out noise due to EMI and other sources, we have incorporated (at room temperature) two Faraday enclosures and low pass filters (BLP −100 MHz mini-circuits) with a pass band of DC to 98 MHz in the measurement circuit. Faraday enclosures are used to intercept the measurement cables from the room temperature measurement equipment and the measurement cable (metal shielded twisted pairs of wires) coming from the top of the cryostat. The first Faraday enclosure is meant to cut off electric field dominated noise pick up and the second enclosure with a magnetic coating is meant to reduce the EMI due to low frequency signals. The transport measurements were done at current of 100 nA, derived from the oscillator output of the lock-in amplifier using a 1 MΩ series resistor.

**Conflicts of interests**

There are no conflicts to declare.

**Acknowledgements**

**We acknowledge** funding from National Physical Laboratory, India and DST, India through the project SR/S2/PU-0003/2010(G), Dilution Refrigerator facility at CSIR-NPL India. A.P. and R.Y. acknowledge CSIR-Junior Research Fellowship (NET).

**References**

1   Hasan, M. Z. & Kane, C. L. Colloquium: topological insulators. *Rev. Mod. Phys.* **82**, 3045 (2010).
2   Moore, J. E. The birth of topological insulators. *Nature* **464**, 194 (2010).
3   Steinberg, H., Laloë, J.-B., Fatemi, V., Moodera, J. S. & Jarillo-Herrero, P. Electrically tunable surface-to-bulk coherent coupling in topological insulator thin films. *Phy Rev B* **84**, 233101 (2011).
4   Chen, J. *et al.* Tunable surface conductivity in $Bi_2Se_3$ revealed in diffusive electron transport. *Phy Rev B* **83**, 241304 (2011).
5   Liu, W. E., Hankiewicz, E. M. & Culcer, D. Weak localization and antilocalization in topological materials with impurity spin-orbit interactions. *Materials* **10**, 807 (2017).
6   Dai, X., Du, Z. & Lu, H.-Z. Negative magnetoresistance without chiral anomaly in topological insulators. *Phy Rev Lett* **119**, 166601 (2017).
7   Breunig, O. *et al.* Gigantic negative magnetoresistance in the bulk of a disordered topological insulator. *Nat Commun* **8**, 15545 (2017).
8   Liu, M. *et al.* Crossover between weak antilocalization and weak localization in a magnetically doped topological insulator. *Phy Rev Lett* **108**, 036805 (2012).
9   Wiedmann, S. *et al.* Anisotropic and strong negative magnetoresistance in the three-dimensional topological insulator $Bi_2Se_3$. *Phy Rev B* **94**, 081302 (2016).
10  Bhattacharyya, B. *et al.* Spin-dependent scattering induced negative magnetoresistance in topological insulator $Bi_2Te_3$ nanowires. *Sci Rep* **9**, 7836 (2019).

11  Banerjee, K. *et al.* Defect-induced negative magnetoresistance and surface state robustness in the topological insulator BiSbTeSe$_2$. *Phy Rev B* **90**, 235427 (2014).

12  Chiu, S.-P. & Lin, J.-J. Weak antilocalization in topological insulator Bi$_2$Te$_3$ microflakes. *Phy Rev B* **87**, 035122 (2013).

13  Li, H. *et al.* Quantitative Analysis of Weak Antilocalization Effect of Topological Surface States in Topological Insulator BiSbTeSe2. *Nano Lett* **19**, 2450-2455 (2019).

14  Shrestha, K. *et al.* Extremely large nonsaturating magnetoresistance and ultrahigh mobility due to topological surface states in the metallic Bi$_2$Te$_3$ topological insulator. *Phy Rev B* **95**, 195113 (2017).

15  Li, H. *et al.* Negative magnetoresistance in dirac semimetal Cd$_3$As$_2$. *Nat Commun* **7**, 1-7 (2016).

16  Zhang, C.-L. *et al.* Signatures of the Adler–Bell–Jackiw chiral anomaly in a Weyl fermion semimetal. *Nat Commun* **7**, 1-9 (2016).

17  Liao, J. *et al.* Observation of Anderson localization in ultrathin films of three-dimensional topological insulators. *Phy Rev Lett* **114**, 216601 (2015).

18  Li, M., Wang, Z., Yang, L., Gao, X. P. & Zhang, Z. From linear magnetoresistance to parabolic magnetoresistance in Cu and Cr-doped topological insulator Bi$_2$Se$_3$ films. *J Phys Chem Solids* **128**, 331-336 (2019).

19  Lang, M. *et al.* Competing weak localization and weak antilocalization in ultrathin topological insulators. *Nano Lett* **13**, 48-53 (2012).

20  Gehring, P., Gao, B. F., Burghard, M. & Kern, K. Growth of high-mobility Bi$_2$Te$_2$Se nanoplatelets on hBN sheets by van der Waals epitaxy. *Nano Lett* **12**, 5137-5142 (2012).

21  Kong, D. *et al.* Few-layer nanoplates of Bi$_2$Se$_3$ and Bi$_2$Te$_3$ with highly tunable chemical potential. *Nano Lett* **10**, 2245-2250 (2010).

22  Bhattacharyya, B. *et al.* Observation of quantum oscillations in FIB fabricated nanowires of topological insulator (Bi$_2$Se$_3$). *J Phy Condens Matter* **29**, 115602 (2017).

23  Sharma, A., Bhattacharyya, B., Srivastava, A., Senguttuvan, T. & Husale, S. High performance broadband photodetector using fabricated nanowires of bismuth selenide. *Sci Rep* **6**, 19138 (2016).

24  Bhattacharyya, B., Sharma, A., Awana, V., Senguttuvan, T. & Husale, S. FIB synthesis of Bi$_2$Se$_3$ 1D nanowires demonstrating the co-existence of Shubnikov–de Haas oscillations and linear magnetoresistance. *J Phys Condens Matter* **29**, 07LT01 (2016).

25  Friedensen, S. E., Parkin, W. M., Mlack, J. T. & Drndić, M. Transmission Electron Microscope Nanosculpting of Topological Insulator Bismuth Selenide. *ACS nano* **12**, 6949-6955 (2018).

26  Friedensen, S., Mlack, J. T. & Drndić, M. Materials analysis and focused ion beam nanofabrication of topological insulator Bi$_2$Se$_3$. *Sci Rep* **7**, 13466 (2017).

27  Assaf, B. A. *et al.* Linear magnetoresistance in topological insulator thin films: Quantum phase coherence effects at high temperatures. *Appl Phy Lett* **102**, 012102 (2013).

28  Zhang, H. *et al.* Weak localization bulk state in a topological insulator Bi$_2$Te$_3$ film. *Phy Rev B* **86**, 075102 (2012).

29  Wang, W. J., Gao, K. H. & Li, Z. Q. Thickness-dependent transport channels in topological insulator Bi$_2$Se$_3$ thin films grown by magnetron sputtering. *Sci Rep* **6**, 25291 (2016).

30  Chiatti, O. *et al.* 2D layered transport properties from topological insulator Bi$_2$Se$_3$ single crystals and micro flakes. *Sci Rep* **6**, 27483 (2016).

31  Shekhar, C. *et al.* Evidence of surface transport and weak antilocalization in a single crystal of the Bi$_2$Te$_2$Se topological insulator. *Phy Rev B* **90**, 165140 (2014).

32  Peng, H. *et al.* Aharonov–Bohm interference in topological insulator nanoribbons. *Nat Mater* **9**, 225 (2010).

33  Chen, J. *et al.* Gate-voltage control of chemical potential and weak antilocalization in Bi$_2$Se$_3$. *Phy Rev Lett* **105**, 176602 (2010).

34  Bansal, N., Kim, Y. S., Brahlek, M., Edrey, E. & Oh, S. Thickness-independent transport channels in topological insulator Bi$_2$Se$_3$ thin films. *Phy Rev Lett* **109**, 116804 (2012).

35  Dey, R. *et al.* Strong spin-orbit coupling and Zeeman spin splitting in angle dependent magnetoresistance of $Bi_2Te_3$. *Appl Phy Lett* **104**, 223111 (2014).

36  Cha, J. J. *et al.* Weak Antilocalization in $Bi_2(Se_xTe_{1-x})_3$ Nanoribbons and Nanoplates. *Nano Lett* **12**, 1107-1111 (2012).

37  Zhang, L. *et al.* Weak localization effects as evidence for bulk quantization in $Bi_2Se_3$ thin films. *Phy Rev B* **88**, 121103 (2013).

38  Wang, Z., Wei, L., Li, M., Zhang, Z. & Gao, X. P. Magnetic field modulated weak localization and antilocalization state in $Bi_2(Te_xSe_{1-x})_3$ films. *Phys Status Solidi (b)* **255**, 1800272 (2018).

39  Wray, L. A. *et al.* A topological insulator surface under strong Coulomb, magnetic and disorder perturbations. *Nat Phys* **7**, 32-37 (2011).

40  Wang, H. *et al.* Crossover between weak antilocalization and weak localization of bulk states in ultrathin $Bi_2Se_3$ films. *Sci Rep* **4**, 5817 (2014).

41  Berry, M. V. Quantal phase factors accompanying adiabatic changes. *P Roy Soc A-Math Phy* **392**, 45-57 (1984).